\documentclass[journal]{IEEEtran}
\usepackage{cite}

\ifCLASSINFOpdf
\usepackage[pdftex]{graphicx}
\else
\fi
\usepackage{amsmath}
\usepackage{url}

\usepackage{balance}
\usepackage{draftwatermark}
\SetWatermarkText{Draft}
\SetWatermarkScale{2}

\begin{document}
%
\title{Users' Concern for Privacy in\\ Context-Aware Reasoning Systems}
%
%
%

\author{Matthias Forstmann,
Alberto Giaretta, and Jennifer Renoux
\thanks{M. Forstmann is with Yale University, New Haven, USA, and with University of Cologne, Cologne, Germany. Email: matthias.forstmann@mail.com}
\thanks{A. Giaretta and J. Renoux are with \"{O}rebro University, \"{O}rebro, Sweden. Email: alberto.giaretta@oru.se , jennifer.renoux@oru.se}
\thanks{This work is supported by the distributed environment Ecare@Home funded by the Swedish Knowledge Foundation, 2015-2019.}
\thanks{Manuscript received June 23, 2020; revised June 23, 2020.}}

%
%

\markboth{Journal of \LaTeX\ Class Files,~Vol.~14, No.~8, June~2020}%
{Shell \MakeLowercase{\textit{et al.}}: Bare Demo of IEEEtran.cls for IEEE Journals}
%



\maketitle

\begin{abstract}
Context-aware reasoning systems allow drawing sophisticated inferences about users' behaviour and physiological condition, by aggregating data from seemingly unrelated sources. We conducted a general population online survey to evaluate users' concern about the privacy of data gathered by these systems. We found that people are more concerned about third parties accessing data gathered by environmental sensors as compared to physiological sensors. Participants also indicated greater concern about unfamiliar third parties (e.g., private companies) as opposed to familiar third parties (e.g., relatives). We further found that these concerns are predicted and (to a lesser degree) causally affected by people's beliefs about how much can be inferred from these types of data, as well as by their background in computer science.
\end{abstract}

\begin{IEEEkeywords}
internet of things, privacy, trust, context aware systems
\end{IEEEkeywords}

%
\IEEEpeerreviewmaketitle

\section{Introduction}
Just a decade ago, the Internet of Things (IoT) was considered a utopian, life-changing technology. The IoT would have contributed to the vision of a future full of interconnected appliances and sensors, making our daily lives easier and more convenient than ever. Fast-forward to today, that vision became a reality. Our lives are permeated with smart devices, ranging from wearable sensors such as smart watches, to entire smart houses riddled with motion sensors, pressure sensors, and even body sensors for real-time health monitoring. Yet, the more pervasive new technologies become, the more privacy and security issues arise.

Modern context-aware systems not only can sense and adapt to their environment, but also are capable of inferring higher-order information from aggregate sensor data~\cite{Wu_2003}. In other words, while individual sensor data may not be sufficient to determine which specific activity a user is engaging in, the combination of multiple sources of data may allow drawing more sophisticated inferences. For example, by combining data from bed pressure sensors and a heart rate monitor (sensors types that individually produce comparably basic data), it is theoretically possible to infer that a user is exercising or having sex. While the average user may be not concerned by the former kind of data leakage, they may well be when the latter is. Yet, we suggest that most users are not fully aware of these inference possibilities and would otherwise adopt a different attitude towards such systems.

In this article, we present the results of an empirical study, conducted to determine how knowledge about the possibility of inferring higher-order information from basic sensor data predicts potential users' concern about who accesses what kind of personal data. In addition, we investigated whether indirect knowledge about data security (using knowledge about computers as a proxy) affects these concerns.

\subsection{Context-Aware Systems and the Internet of Things}
Current context-aware systems typically use pervasive computing to capture information about the users' environment and infer activities such as cooking, sleeping, working out, or leaving the house~\cite{krishnan2014activity}. The addition of wearable sensors to these activity recognition systems, especially in the form of smart phones and smart watches, makes it possible to increase the range of recognisable activities, for example in the domains of personal well-being or outdoor activities~\cite{incel2013review}. However, context recognition is more than just recognising activities: it can similarly be applied to monitor health-related parameters or to recognise emergency events. As more body sensors become available, introducing them in context-aware systems enables low-cost, real-time, and non-invasive monitoring of users' physiological condition~\cite{alirezaie2017ontology}. 

Despite these advantages, the introduction of a large number of sensors in people's lives poses a threat to their privacy. By design, in an IoT system, multiple types of devices (and therefore their sensors) are directly or indirectly connected and able to interact. Consequently, due to the fact that sensitive information can be inferred by combining seemingly unrelated data, the data collected needs to be handled in a secure manner and users' privacy guaranteed. Concerns about privacy aspects of context-aware systems are indeed not unwarranted: the sudden rise of demand for IoT devices lead companies to spend less time on research and development, and to introduce their products to the market as quickly as possible. In turn, since the average customer does not perceive security as an added value, other aspects of the product were prioritised. As a result, a considerable number of products were shipped with critical security flaws, from tea kettles to insulin pumps, thermostats, baby monitors, home alarm systems, and X-Ray systems~\cite{DraGiaMaz17}. Eventually, malicious actors started exploiting these vulnerabilities, consolidating IoT devices in botnets and delivering powerful DDoS attacks, such as the ones performed by the “Mirai” malware in 2016~\cite{dedonno2017analysis}.

When deployed within the setup of a smart home, a context-aware system might thus expose users to a number of privacy-related risks. Malicious attackers might break into one, or multiple, context-aware systems to steal sensitive data that have already been processed by the respective device (that is, accessing information which the device was designed to provide). Intruders might also implement their own reasoner in the device, directly accessing sensors in order to independently derive information about the victim that go beyond the intended purpose of the device. While the former may be the most straightforward approach, it is not necessarily the easiest nor most profitable. Since the reasoner is one of the most critical elements of a context-aware system, most effort is typically spent securing this part of the system, while the IoT sensors themselves are often left less protected, making them an easier target for malicious third parties. As a result, an attacker may be able to gather personal information about the victim. Information that the victim does not even now it can be inferred from the available sensors. Yet, why does it matter whether or not an actor infers things about a user from basic sensor data?

\subsection{Trust and Perception of Privacy in IoT Systems}
People fundamentally care about what others know about them, and strategically control which information is available to which social entity. Behind this motivation stands a desire to have others think highly of us—to think of us as honest, moral, and friendly people (\textit{impression management}~\cite{leary1990impression}). Such an external assessment is a prerequisite to fulfil our basic “need to belong”, that is, our natural tendency to want to be an accepted part of a social group~\cite{baumeister1995}. As such, we take measures to assure that information that may be detrimental to this cause does not reach the public—that only we can access it. This similarly applies to the digital world: it has been argued that concern for privacy, for example in instant messaging applications, is due in part to people’s desire to manage others’ impression of them~\cite{kobsa2012}.

In addition, access to sensitive information about us, or to exclusive knowledge that we may have, can enable potentially malicious third parties to exploit us, be it socially (by threatening our social status), financially (by gaining access to our resources), or even physically. Accordingly, research shows that people indeed care about the privacy of their data~\cite{Belanger2011}, for instance in the context of e-government~\cite{Abri2009}, e-commerce~\cite{Tsai2011} or the use of RFID chips in products~\cite{Cazier2008}. What has been established thus far is that concern for privacy (defined by Westin (1967) as “the claim of individuals […] to determine for themselves when, how, and to what extent information about them is communicated to others” (p. 7)) has actual consequences for people's behaviour, for example in determining whether or not they want to use a given system~\cite{BrownMuchira2004}.

\subsection{User's Trust in IoT Devices}
Interestingly, however, while people overtly claim to be extraordinarily concerned about digital privacy, they seem substantially less concerned about using everyday technologies that can collect and process personal information~\cite{Nguyen2008}. They seem to readily trust devices that can access private data, but why?

As a fundamental building block of human cooperation, people generally tend to trust one another. In other words, we intuitively think of most other people as sincere, benevolent in their intentions, and truthful in their communication~\cite{acar2014trusting,wrightsman1991interpersonal,mayo2015}. This social orientation provides us with an essential sense of safety and security~\cite{schul2008value}, but carries with it the risk of being exploited. In fact, people are aware that others may try to mislead them, and they only trust them for as long as they don’t give them reasons not to~\cite{berg1995trust}.

Distrusting others requires more cognitive resources, because people have to engage in, for example, constant monitoring and assessment of the distrusted party~\cite{mayo2015}. However, it can be beneficial to detect malicious actors before they can harm us. Whether we tend to distrust or trust another person depends on certain personality variables~\cite{mayo2015}, but also on situational parameters. Specifically, depending on environmental variables (for example, how much is at stake), a person must decide whether or not trusting a third party is worth the risk~\cite{mayer1995integrative}. As a consequence, if people are not entirely in the know about these situational parameters, they may wrongfully trust a third party too readily. In the present case, not knowing about the inferences that can be drawn from their personal data may thus correspond to higher levels of trust towards context-aware systems. In fact, research suggests that this privacy paradox (i.e., being concerned about privacy while still using privacy-threatening devices) occurs in the IoT domain, precisely because people do not realise how knowledge about individual activities can be aggregated to infer higher-order information, resulting in a disregard for privacy protection~\cite{williams2017privacy}. According to some theorising~\cite{nehf2003}, this disregard is not just an end-user problem, but a problem for society as a whole.

A recent technical report~\cite{groopman2015consumer} sheds more light on IoT users' privacy concerns. While they indicate considerable discomfort at the thought of their data being sold to companies, the majority of them admit that they do not fully understand their devices or basic privacy threats. They admit to not knowing how companies use their data, and only marginally care for this information. Additionally, participants were proposed seven different categories of IoT environments (e.g., their home, private transportation, public market places) and, when asked about how important is to them to be notified when data is collected, they rated their own home and their own bodies as the most critical environments (see also Bahirat et al.~\cite{bahirat2018data} and Lee and Kobsa~\cite{lee2017privacy} for related investigations). In this study, we focus on sensors used in those two environments.

To the best of our knowledge, there has to date neither been an extensive investigation of users' understanding of the quantity of information that can be inferred from their personal data, nor into how beliefs or knowledge about such inferability affects their concern for privacy. Therefore, we designed the present study to investigate if and how knowledge about inferability of higher order information from basic IoT sensors can predict users' concern for privacy, and whether this concern depends on the type of sensor data accessed and/or the familiarity of the party potentially accessing this data.

\section{The Present Research}
In the present research, our goal was to answer a set of questions about user’s concern regarding the privacy of data stemming from context-aware reasoning systems. In a first step, we sought to assess people’s general level of concern about different types of IoT devices (1). Subsequently, we tested the hypothesis that greater knowledge about how much can be inferred from sensor data predicted concern about context-aware systems. This was implemented twofold. First, we assessed participants’ general belief about how much can be inferred from these data and tested how predictive this would be of concern about others accessing their private data (2a). Second, we tried to test whether strengthening participants’ knowledge about the inferability of these data would increase concern (2b). We then tested whether participants differentiate in their concern between health-related and environmental sensor data (3), and whether familiarity of parties potentially accessing the different types of sensor data established in the previous step would affect participants’ concern (4). Last, we exploratorily analysed how general knowledge about data security (approximated via general computer knowledge) related to concerns about different parties’ access to private data (5).

\subsection{Participants and Design}
A total of 304 participants were recruited via the online crowd-sourcing platform Amazon Mechanical Turk (MTurk), in exchange for a modest monetary compensation. We refer the reader to Buhrmester et al., for a discussion on MTurk data quality~\cite{buhrmester2011amazon}. Of those participants, 17 were excluded from data analyses as they indicated on a single binary item that they responded to one or multiple questions in a random (or purposefully wrong) manner, leaving a final sample of 287 participants (152 female, 134 male, 1 other; $M_{Age} = 40.83, SD = 12.07$). In addition to assessing their person beliefs about inferability, participants were randomly assigned to one of two conditions, either providing them with additional information about inferences that could be drawn from sensor data (\textit{info} condition) or not (\textit{no-info} condition). 

\subsection{Materials and Procedure}
Upon consent, all participants were successively introduced to 13 types of sensors, comprising environmental, biometric, and physiological types (see Figure~\ref{fig:concern_sensors} for the complete list). In the no-info condition, participants were only given little information about what each sensor measures (e.g., "Temperature and humidity sensors are currently used to adjust the level of a home thermostat and humidifier."), while participants in the info condition received additional information about what kind of inferences could be drawn by combining these data with other (e.g., "Existing systems are able to infer that you are showering or taking a bath, by using the level of humidity and the temperature in your bathroom as indicators. Combined with other indicators […] and in other locations it can also be used to infer if you are toileting, doing your laundry or cooking.")

For each sensor, participants in both conditions were asked to indicate on a single 7-point Likert-type scale ranging from 1 (not at all) to 7 (very much) how concerned they would be about using such a sensor (general concern). In addition, using the same scale, participants were asked for each sensor how concerned they would be if the monitored values could be accessed by them (exclusively), their relatives, their doctors, a private organisation, and a public organisation, respectively.

Subsequently, as our main predictor variable, participants answered two items assessing their general belief in how much can be inferred from sensor data. The two items read "I think it is possible to infer a great number of things about my life/my behaviour from seemingly unrelated data" and "Sometimes, very elaborate things can be inferred from combining simple measurements". Participants were asked to indicate their agreement on scales ranging from 1 (disagree completely) to 9 (agree completely). Both items were highly correlated ($r(287) = .814, p < .001$) and therefore averaged into a single score representing participants’ beliefs about inferability.

To conclude, participants provided basic demographic information (age, gender, nationality, level of education, etc.), indicated whether they had an educational or work-related background in computer science (yes/no), and answered on a 7-point scale how much they generally knew about computer science (very little to very much). Finally, participants responded to the binary exclusion item referenced above.

\subsection{Data reduction} \label{sec:data_reduction}
For ease of data interpretation, we employed two data reduction strategies, allowing us to create mean scores across parties potentially accessing the data as well as the different sensor types.

\subsubsection{Data Access} We assumed a-priori that participants' responses regarding their concern about who can access their data would be explained by three underlying factors: access by themselves, access by familiar others (relatives/doctors) and access by unfamiliar/untrusted others (private/public organisations). The “access by themselves” category allows us to assess how much users are generally receptive to the idea of having certain IoT, without considering possible leaks or usage of data by third parties. The “access by familiar others” and “access by unfamiliar others” categories, conversely, allow us to consider the degree of concern users have about other’s having access to their data. 

To test this model, we ran a confirmatory factor analysis~\cite{kline2015principles} using the R-package lavaan~\cite{rosseel2012lavaan}. The model produced a Comparative Fit Index (CFI) of $0.971$ and a Standardized Root Mean Square Residual (SRMR) of $0.046$, indicating good fit to the data~\cite{hu1999cutoff}. Standardized factor loadings were in line with expectations for own access (fixed to $1.0$ on factor 1), access by relatives and doctors ($.819$ and $.913$ on factor 2), as well as for access by private and public companies ($.958$ and $.994$ on factor 3). Concern about access by familiar others covaried positively with concern about own access ($\beta = .377, SE = .058, p < .001$) and even more so with concern about unfamiliar others' access ($\beta = .693, SE = .036, p < .001$), whereas concern about own access and unfamiliar others' access covaried marginally significant and negatively ($\beta = -.107, SE = .060, p = .073$).

Therefore, for all types of sensor data, we created mean scores for concern about access by familiar and unfamiliar others, and will primarily use those (in addition to the "access by self" score) in the remaining of this article.

\subsubsection{Sensor Types}
Using a similar approach, we tested whether access concerns regarding the 13 different sensors could be explained by a limited set of underlying factors. As we did not have a-priori hypotheses about the factor structure, we ran an exploratory principal-component factor analysis across the aggregate concern-scores for the different sensor types, using Oblim rotation with Kaiser normalization~\cite{fabrigar1999evaluating}.

The analysis produced two factors with Eigenvalues greater than 1, explaining a combined 79\% of the total variance. The first factor comprised all physiological sensors (all factor loadings \textgreater .61) including keystroke dynamics, while the second factor comprised the five environmental sensors (all factor loadings \textgreater .68). Concern about access to fingerprint sensors loaded equally weak on both factors (with factor loadings of .407 and .422 respectively).

As such, it seems as if people mentally cluster their concern about access to sensor data. They have distinguishable concern about access to environmental versus physiological data. We therefore created two additional aggregate scores representing these factors. Concern about fingerprint sensors was not included in either score. In summary, we calculated the following aggregate scores:

\begin{enumerate}
	\item \textbf{General concern} about sensors generating data (i.e., how concerned users are about using the sensors in general).
    \item Concern about \textbf{accessing party} (i.e., how concerned users are about the data being accessible to a specific entity). We distinguish three types of entities: \textit{oneself}, \textit{familiar others} (relatives and doctors) and \textit{unfamiliar others} (public and private organizations).
    \item Concern about access to \textbf{data types} (i.e., how concerned users are about a specific type of data being monitored and accessed). We distinguish two types of data: environmental (humidity, temperature, motion, pressure and door states) and physiological (blood sugar, blood pressure, grip, heart rate, skin reactivity, gait and keystroke dynamics).
\end{enumerate}

\section{Results}\label{sec:results}
In the following, we will first discuss participants' general concern about the use of IoT sensors and the accessibility of different types of data by different parties, analysed across all participants. We will then focus on how beliefs about inferability relate to data access concerns, additionally considering the effect of the experimental manipulation (info vs. no-info condition) on these concerns. Finally, we will exploratorily analyse the role that users' computer science background play in their concern about data access.

\subsection{General Concern (Descriptive)} \label{subsec:general}
Although the scale we used to assess this construct did not have an objectively neutral midpoint, we decided to descriptively analyse participants’ responses by testing their scores against the scale midpoint, in order to get a rough estimate of whether or not they tend to be concerned about using any or all of the sensors. Across all types of sensors, participants revealed a level of general concern that was significantly lower than the scale midpoint: $M = 3.48$, $SD = 1.52$, $t(286) = -5.77$, $p < .001$, $d = -0.34$. In fact, looking at the individual sensors, participants only indicated general concern higher than the scale midpoint for fingerprint data ($M = 4.79$, $SD = 2.02$, $t(286) = 6.62$, $p < .001$, $d = 0.39$), and -- albeit not significantly -- for keystroke data ($M = 4.19$, $SD = 2.25$, $t(286) = 1.43$, $p = .152$, $d = 0.08$), while all other scores averaged (sometimes marginally) significantly below the midpoint, except for pressure data, which did not significantly differ from the scale midpoint ($p = .24$).

In other words, participants were not overly concerned about the possibility of having these sensors in their homes or attached to their bodies, except for sensors gathering biometric data. This high degree of concern may be due to the common use of biometric sensors as means for personal identifications and concern about either unwanted identification by untrusted third parties or identity theft. In addition, based on the results of the factor-analysis outlined above, we found that participants were in general not significantly more concerned about having environmental sensors in their homes ($M = 3.39$, $SD = 1.59$) than they were about having physiological sensors attached to their bodies ($M = 3.36$, $SD = 1.76$, $t(286) = -0.44$, $p = .662$, $d = -0.02$).

\subsection{Data Access Concern} \label{subsec:access}
\begin{figure}[t]
\includegraphics[width=0.45\textwidth]{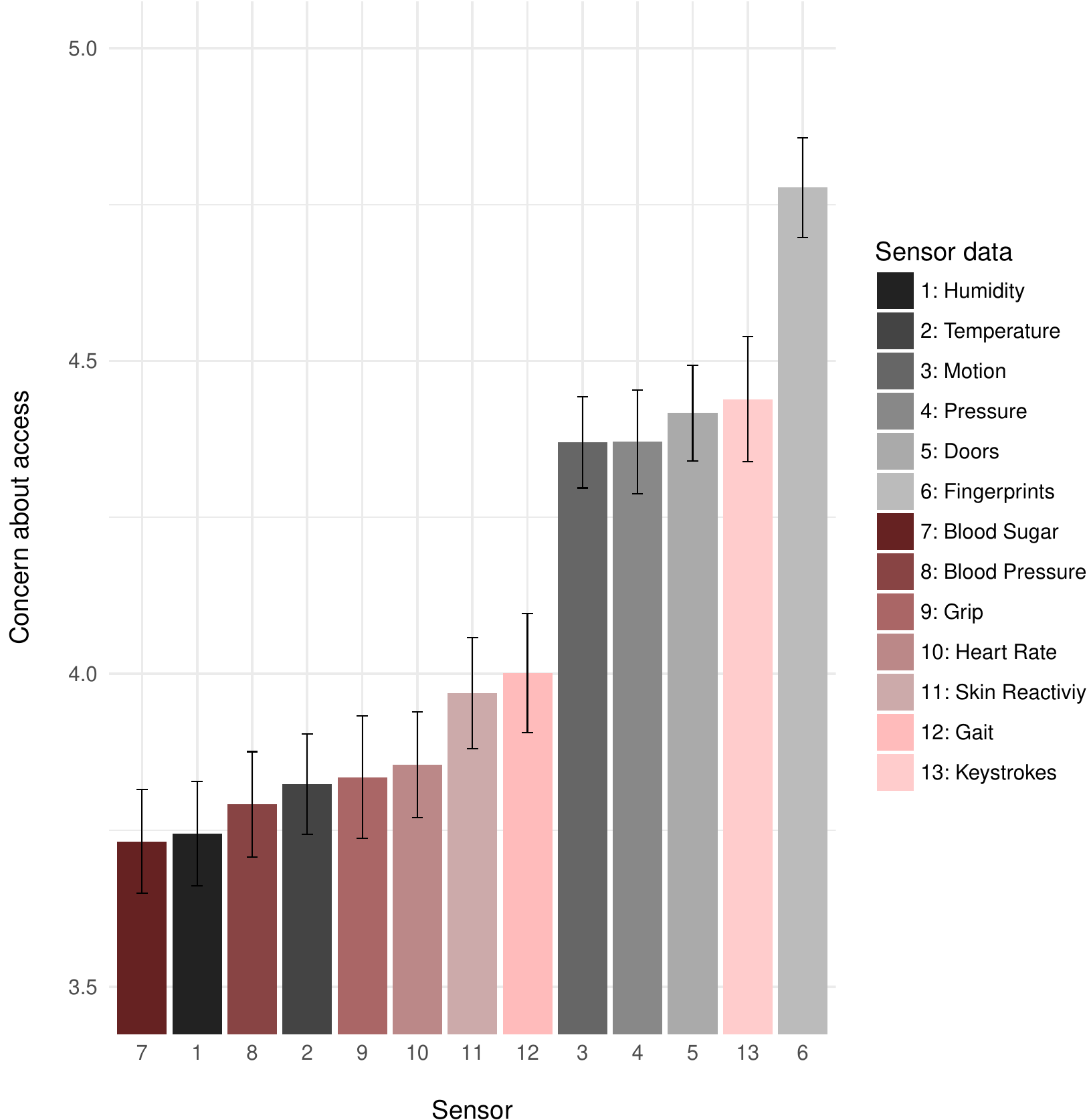}
\caption{Mean concern expressed by participants with respect to each sensor. Error bars indicate standard errors of the mean.}
\label{fig:concern_sensors}
\end{figure}

Results are different, however, when looking at people's concern about who can access their data. Across all types of sensors, people's concern about who has access to their data was higher ($M = 4.09$, $SD = 1.21$) than their general concern about having the sensors at home or attached to their bodies ($M = 3.48$, $SD = 1.52$, $t(286) = 7.20$, $p < .001$, $d = 0.43$). This indicates that people do not care as much about the presence of the sensors per se, but rather about who can access the measurements. Regarding the latter, people show a clear pattern favouring familiarity. They are less concerned when they have exclusive access to the data ($M = 1.82$, $SD = 1.29$) than when familiar others have access ($M = 3.67$, $SD = 1.62$, $t(286) = -19.23$, $p < .001$, $d = -1.26$), and less concerned about familiar others than unfamiliar others ($M = 5.64$, $SD = 1.59$, $t(286) = -23.12$, $p < .001$, $d = -1.22$). Similarly, participants clearly distinguish between which kind of data is accessed (see Figure~\ref{fig:concern_sensors}). Across all accessing parties, they were more concerned about who has access to environmental measures ($M = 4.15$, $SD = 1.21$) than to physiological measures ($M = 3.95$, $SD = 1.37$, $t(286) = 3.86$, $p < .001$, $d = 0.15$).

\begin{figure}[t]
\begin{center}
\includegraphics[width=0.45\textwidth]{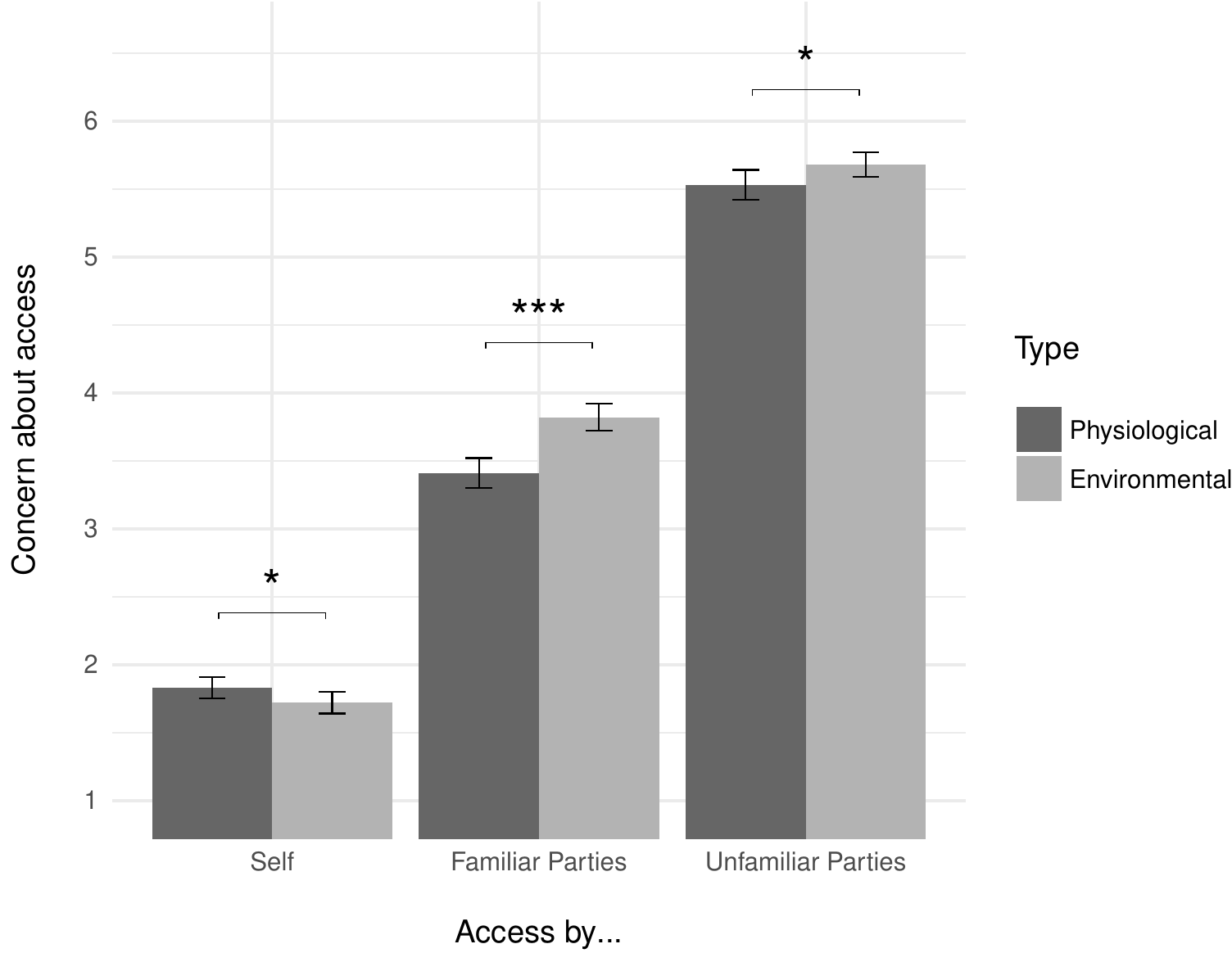}
\caption{Relation between access concerns and familiarity with the accessing third party by data type. Asterisks indicate level of significance: * = $p < .05$, ** = $p < .01$, *** = $p < .001$ }
\label{fig:physiological_vs_environmental} 
\end{center}
\end{figure}

To get at a more detailed picture of this pattern, we additionally ran a 2 (sensor type: environmental vs. physical) x 3 (accessing party: self vs. familiar vs. unfamiliar) within-subjects ANOVA that produced a significant interaction effect ($F(2, 286) = 26.45$, $p <.001$, $\eta_p^2 = .085$), displayed in Figure~\ref{fig:physiological_vs_environmental}. In addition to the two main effects detailed above, the individual contrasts reveal that when people have exclusive access to their data, they tend to be more concerned about physiological data ($M = 1.83$, $SD = 1.37$) than about environmental data ($M = 1.72$, $SD = 1.28$), $t(286) = 2.44$, $p = .015$, $d = 0.09$).

This pattern flips, however, when data are accessed by familiar and unfamiliar others. In these cases, people are more concerned about unfamiliar others having access to their environmental ($M = 5.68$, $SD = 1.56$) than to their physiological measures ($M = 5.53, SD = 1.79, t(286) = 2.37, p = .018, d = 0.08$), and this applies even more so to familiar others ($M = 3.41, SD = 1.83$ vs. $M = 3.82, SD = 1.68, t(286) = 5.31, p < .001, d = 0.23$). Correspondingly, all three individual 2 x 2 interactions turned out statistically significant (for all: $p < .001$ and $\eta_p^2 > .044$).

We suggest two explanations for this pattern of results. First, we believe that people are concerned about others knowing about their private behaviour, which can be inferred from environmental measures (e.g., who is in their house and what activities they engage in). Especially familiar others could directly confront them with certain (potentially embarrassing) behaviours inferred from the sensor data. Second, the potential benefit of enabling familiar others to access their physiological data (e.g., to react appropriately in case of a medical emergency) might account for the lower levels in users' concern about these parties accessing this type of data. Why participants tend to be more concerned about the use of physiological sensors than environmental sensors when they themselves have exclusive access to the data (although concern for both is relatively low overall) remains an open question. One possible explanation could be that people perceive being confronted with their vital signals as a stressor or constant reminder of unhealthy habits, leading to a slight preference for the use of environmental sensors.

\subsection{Knowledge About Possible Inferences}\label{subsec:info}
To answer our main question, we investigated how knowledge/beliefs about how much can be inferred from sensor data predicts usage and data access concerns. We found that while stronger belief in how many inferences can be drawn from measurement data did not predict general concern about having the sensors at home or attached to one's body ($\beta = .078, SE = .059, p = .190$), it significantly predicted concerns about who can access the data ($\beta = .205, SE = .058, p < .001$).

\begin{figure}[t]
\begin{center}
\includegraphics[clip, width=0.45\textwidth]{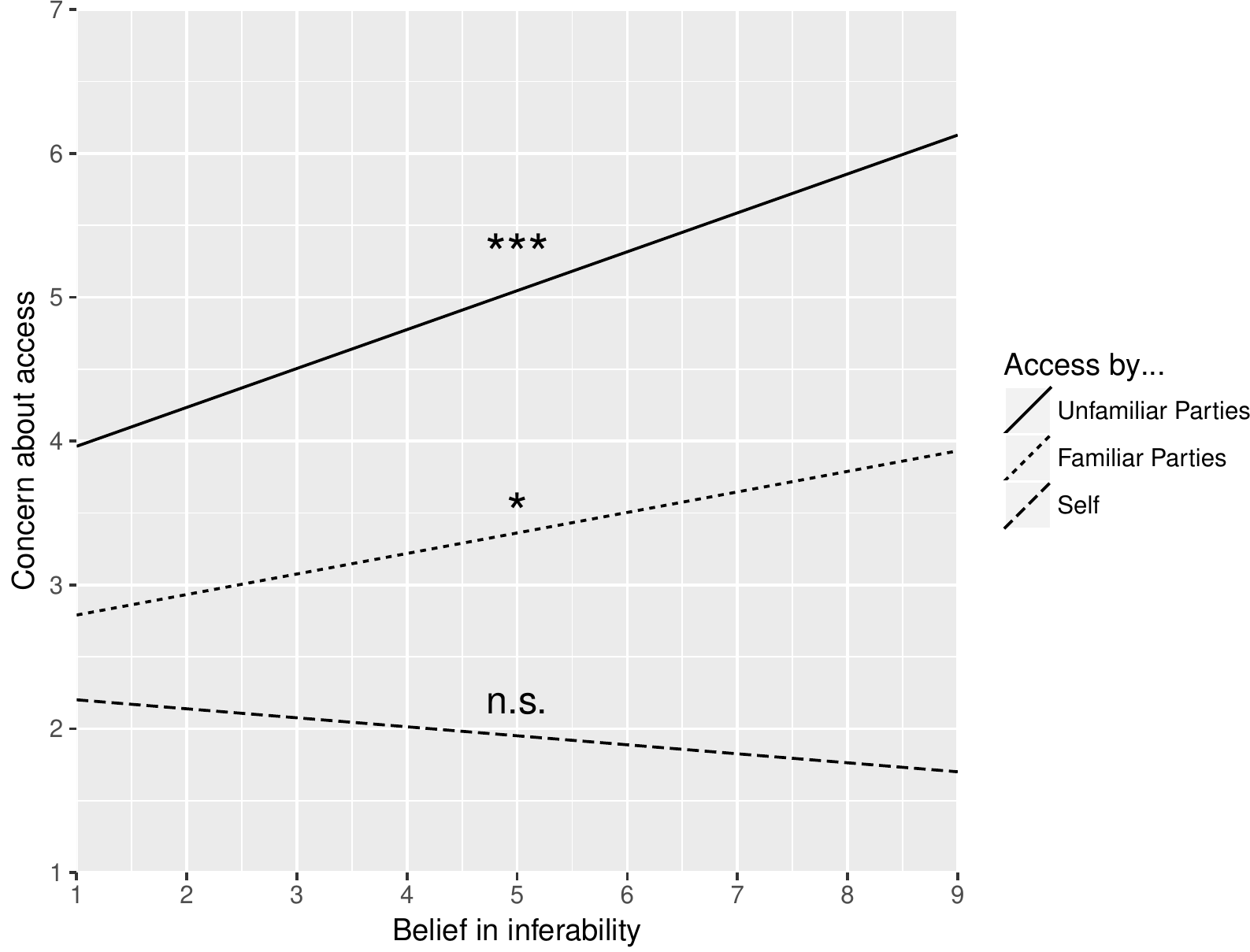}
\caption{Relation between data access concerns and knowledge/beliefs about inferability.}
\label{fig:concern_unfamiliarity} 
\end{center}
\end{figure}

In order to control for intercorrelation between the dependent variables, we used a structural equation model (SEM) to test how much participants' belief uniquely predicted concern about different parties potentially accessing their data (across all sensor types). We found that a stronger belief in inferability does not predict concern about data that only oneself has access to ($\beta = -.079, SE = .059, p = .177$), but it does significantly predict concern about familiar others ($\beta = .143, SE = .058, p = .031$), and unfamiliar others ($\beta = .278, SE = .053, p < .001$) having access to the data (Figure~\ref{fig:concern_unfamiliarity}). In both cases, greater belief in inferability corresponded to higher levels of concern.

Comparing the different slopes, belief in how easy inferences can be drawn from measurement data has a stronger predictive effect for concern about familiar others accessing the data than for oneself accessing the data ($\beta_{diff} = .222, SE = .064, p < .001$), and a stronger predictive effect for concern about unfamiliar others than for familiar others ($\beta_{diff} = .134, SE = .051, p = .008$) and oneself ($\beta_{diff} = .357, SE = .083, p < .001$). As such, belief/knowledge about how much can be inferred from sensor data is primarily associated with concern about unfamiliar third parties accessing one’s data. 
As a second SEM revealed, concern for data produced by environmental versus physiological sensors was not differently predicted by participants’ beliefs, $\beta_{diff} = .035, SE = .039, p = .358$.

\subsubsection{Experimental Manipulation}
In addition, all participants were assigned to one of two experimental conditions, in which they either received information about the kind of data that could potentially be inferred from the measurement data (info condition), or they received no additional information (no-info condition). Adding experimental conditions (info vs. no-info) as an additional between-subjects factor to the sensory type $\times$ accessing party ANOVA (reported above) did not produce a significant 2 x 3 x 2 interaction ($F(2, 284) = 0.38$, $p = .680$, $\eta_p^2 = .003$), nor did it significantly interact with either of the other two factors ($p = .817$ and $p = .414$, respectively). In other words, the manipulation did not differentially affect participants' concern about physiological versus environmental data being accessed by them, familiar, and unfamiliar parties.

However, the manipulation produced a significant main effect on people's access concerns ($F(1, 285) = 4.25$, $p = .04$, $\eta_p^2 = .003$), in that receiving additional information about what can be inferred from the measurement data increased people's access concerns across all sensors. Looking at the individual sensors, the effects on humidity ($t(285) = 2.16, p = .031, d = 0.26$), pressure ($t(285) = 2.55, p = .011, d = 0.30$), and keystroke dynamics ($t(258) = 2.17, p = .031, d = 0.26$) were statistically significant, with marginally significant effects for temperature ($t(285) = 1.80, p = .073, d = 0.21$) and heart rate sensors, $t(285) = 1.79, p = .074, d = 0.21$. Across all sensors, the experimental manipulation did not affect people's general concern about having the sensors in their homes or attached to their bodies ($t(285) = 0.94, p = .348, d = 0.11$). Surprisingly, participants’ score reflecting their belief/knowledge about how much can be inferred from sensor data, which we used in our SEMs as predictor variable, was not affected by having had exposure to additional information, $M = 7.17, SD = 1.59$ vs. $M = 7.20, SD = 1.67, t(285) = 0.15 , p = .877, d = 0.02$.

In sum, it seems as if beliefs/knowledge about how inferences can be drawn from data is (causally) positively related to the concern about who has access to such data. This seems to be especially the case for access by unfamiliar third parties.

\subsection{Knowledge About Computer Science and Other Demographics}\label{subsec:background}
\begin{figure}[t]
\begin{center}
\includegraphics[clip, width=0.45\textwidth]{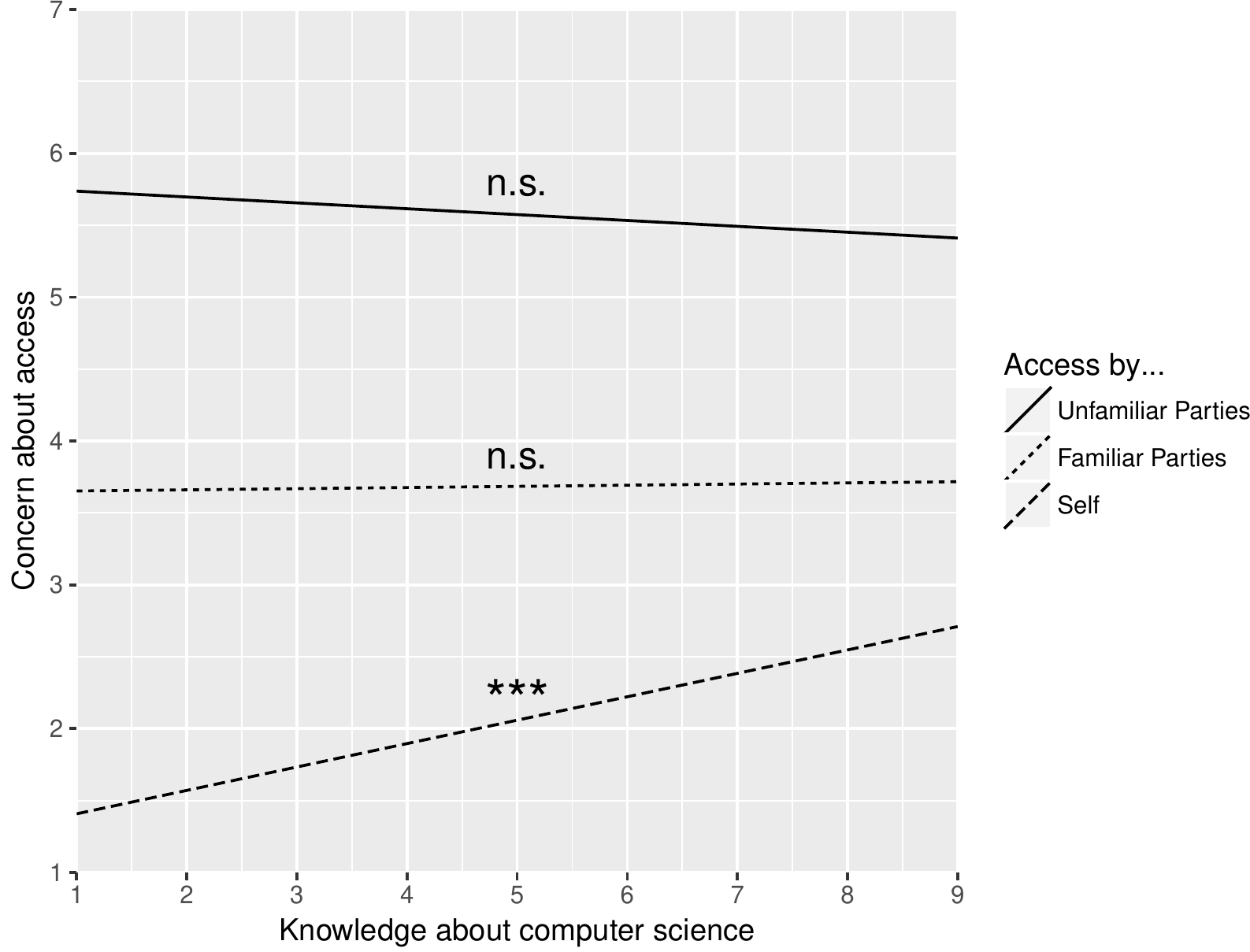}
\caption{Relation between data access concerns and background in computer science.}
\label{fig:concern_cs_background}
\end{center}
\end{figure}

In addition to the analyses detailed above, we also exploratorily analysed whether knowledge about computer science predicts people's concern about data access. While general knowledge about computer science did not correlate with belief in how much can be inferred from measurement data ($r(287) = 018, p = .759$), a SEM revealed that this knowledge positively predicts how concerned people are about data to which only they have access ($\beta = .206, SE = .056, p < .001$), while it did not predict concern about access by familiar ($\beta = .008, SE = .059, p = .891$) or unfamiliar ($\beta = -.042, SE = .059, p = .476$) third parties (see Figure~\ref{fig:concern_cs_background}). Confirming this notion, the same pattern was found for our binary variable asking participants about whether or not they have an educational background in computer science. While people with such a background were more concerned about data to which only they have access ($n = 59, M = 2.13, SD = 1.62$) than those without such background ($M = 1.73, SD = 1.18 t(285) = 2.18, p = .030$), this was not the case for concern about others accessing their data (familiar: $p = .605$, unfamiliar: $p = .793$).

One explanation for these results could be that advanced knowledge about computer science may not necessarily be related to knowledge about inferences that can be drawn from data, but it may be associated with awareness of the fact that data are almost never fully secure, and always potentially accessible to malicious third parties. We further found that knowledge about computer science positively correlated with general concern about wearing the sensors or having them at home ($r(287) = .146, p = .014$). This concern was similarly predicted by the binary variable, with those who have a background in computer science being more concerned ($M = 3.97, SD = 1.57$) than those without ($n = 228, M = 3.27, SD = 1.57, t(285) = 2.82, p = .005, d = 0.33$). This finding could likewise be related to advanced knowledge about general data insecurity that people with a background in computer sciences may have, regardless of inferences that could be drawn from the data. 

Finally, we found that participants' age correlated with belief in inferability ($r(287) = .171, p = .004$), and concern for who accesses their data ($r(287) = .103, p < .082$), with an especially pronounced concern for access by public and private organisations ($r(287) = .176, p = .003$). Age does also negatively correlate with knowledge about computer science ($r(287) = -.120, p = .043$). These findings are in line with general findings about younger people having lesser concern about sharing personal data with unfamiliar third parties~\cite{taraszow2010disclosure}.

\section{Conclusion}
Thus far, little was known about how aware potential users of IoT systems are of the inferences that can be drawn from basic sensor data, and how such an awareness relates to their concerns about using these devices depending on who might access the data produced. In the present article, we presented the results of an online study conducted with a US-American participant pool, investigating users’ concerns about the monitoring of (and others’ access to) environmental and physiological sensor data. The following list sums up the main findings from this study:

\begin{enumerate}
	\item People mentally cluster sensor types into two categories (environmental and physiological sensors), and have different levels of concern about access to the data produced by these two types of sensors. They further cluster accessing parties into three categories: themselves, familiar others (doctors and relatives), and unfamiliar others (public and private companies).
    \item Global concern about using environmental and physiological sensors is generally low and comparable in strength, with the exception of biometric sensors, which are typically used for identity verification and therefore presumably associated with corresponding threats such as identity theft.
    \item Across sensors, users are more concerned about who may potentially access their data than they are about usage of the sensors per se.
    \item Users are more concerned about data being accessible to familiar others than about data being exclusively accessible to themselves. Furthermore, they indicate an even higher level of concern about data being accessible to unfamiliar others.
	\item Users are more concerned about familiar and unfamiliar others accessing their environmental data than their physiological data, presumably because environmental data reveals more about private activities. Yet, determining the exact causes of this effect would require additional studies.
    \item Pertaining to our main research question, users who believe more strongly that significant information can be inferred from environmental and physiological data are concerned to a greater degree about their data being accessed by familiar and (even more strongly) unfamiliar third parties.
    \item Providing users with additional information about how much can be inferred by combining sensor data increases users' concern about data access across sensor types and accessing parties.
    \item Knowledge about computer science predicts people's concern about data that is ostensibly only accessible to them, presumably because such an educational background is associated with knowledge about the inherent insecurity of data storage. However, additional research would be necessary to adequately investigate this relationship.
\end{enumerate}

One limitation of the present study is the fact that all participants were located in the United States, a prototypical W.E.I.R.D. culture (Western, Educated, Industrialized, Rich, and Democratic~\cite{henrich_heine_norenzayan_2010}). Indeed, research has shown that culture and environment play an important role in users' concern for privacy~\cite{bellman2004}. Westin~\cite{westin1968privacy} talks about privacy as ``an instrument for achieving individual goals of self-realization'' (p. 39). Yet, such goals may critically depend on people’s self-construal. For example, whether people define themselves in terms of their group membership (i.e., and interdependent self-construal) - more frequently encountered in Asian cultures - as opposed to their internal attributes (i.e., an independent self-construal), was found to affect their concerns for privacy~\cite{xu2007effects}. Drawing on these findings, it would certainly be a worthwhile research endeavour to replicate the present study within different cultural contexts.

One should further note that the present study was administered with the help of an online data collection service. Despite its convenience for the researcher, such an approach artificially limits participants to those with sufficient knowledge about new technologies to know and use this service. Whether people who lack this knowledge would consider using IoT devices, and whether their privacy concerns diverge from the present findings, can similarly not be answered without further (and more sophisticated) data collection, for example by using more representative samples. However, an exploratory analysis of the present dataset suggests that knowledge about computer science has indeed some effects on people’s privacy concerns.

In addition, although our results point towards the notion that knowledge about inferences affect people’s concerns about who may potentially access their data, our study did not investigate behavioural consequences of these concerns. In other words, it remains unclear whether people are more or less willing to actually use IoT devices, whether they would only acquire them if they come from a trusted source, or whether they would willingly share their data—for example with health professionals. As in the case of the other limitations, future research may address some of these questions.

Lastly, the question remains whether and how people’s minds could be changed when it comes to their concerns for privacy in the IoT context. As the present research shows, even providing them with information about what can be inferred about their personal behaviour and their physiological constitution by combining data from various IoT sensors does not increase their concerns to a degree that would make them considerably more cautious about owning such devices in the future. Further, it is unlikely that information about inferability would be relied to the consumer by the manufactures of these devices. As such, thorough education of people of all ages about the benefits and risks of utilising content aware systems remains a critical endeavour in the digital age—especially considering the growing presence of such systems in homes, smart devices, or even public spaces.
\balance
\bibliographystyle{IEEEtran}
\bibliography{references}

\end{document}